\documentclass[12pt,a4paper,final]{iopart}
\usepackage{iopams}  
\usepackage{graphicx}
\usepackage[breaklinks=true,colorlinks=true,linkcolor=blue,urlcolor=blue,citecolor=blue]{hyperref}
\usepackage[T1]{fontenc}
\usepackage{graphics}
\usepackage{textcomp}
\usepackage{amssymb}
\usepackage{nicefrac}
\usepackage{color}
\usepackage{setspace}
\usepackage{epstopdf}

\begin{document}

\title[Characterization of a 450-km Baseline GPS Carrier-Phase Link]{Characterization of a 450-km Baseline GPS Carrier-Phase Link using an Optical Fiber Link}

\author{Stefan Droste$^{1}$, Christian Grebing$^{2}$, Julia Leute$^{2}$, Sebastian M.F. Raupach$^{2}$, Arthur Matveev$^{1}$, Theodor W. Hänsch$^{1,4}$, Andreas Bauch$^{2}$, Ronald Holzwarth$^{1,3}$ and Gesine Grosche$^{2}$}
\address{$^1$Max-Planck-Institut für Quantenoptik, Hans-Kopfermann-Str. 1, 85748 Garching, Germany}
\address{$^2$Physikalisch-Technische Bundesanstalt, Bundesallee 100, 38116 Braunschweig, Germany}
\address{$^3$Menlo Systems GmbH, Am Klopferspitz 19a, 82152 Martinsried, Germany}
\address{$^4$Ludwig-Maximilians Universität, Schellingstrasse 4, 80799 München, German}
\ead{stefan.droste@nist.gov}

\begin{abstract}
A GPS carrier-phase frequency transfer link along a baseline of 450~km has been established and is characterized by comparing it to a phase-stabilized optical fiber link of 920~km length, established between the two endpoints, the Max-Planck-Institut für Quantenoptik in Garching and the Physikalisch-Technische Bundesanstalt in Braunschweig. The characterization is accomplished by comparing two active hydrogen masers operated at both institutes. The masers serve as local oscillators and cancel out when the double differences are calculated, such that they do not constitute a limitation for the GPS link characterization. We achieve a frequency instability of $3~\times~10^{-13}$ in 30~s and $5~\times~10^{-16}$ for long averaging times. Frequency comparison results obtained via both links show no deviation  larger than the statistical uncertainty of $6~\times~10^{-16}$. These results can also be interpreted as a successful cross-check of the measurement uncertainty of a truly remote end fiber link.
%200-300 words max, no references

\end{abstract}

\vspace{2pc}
\submitto{\NJP}
%\maketitle

%\tableofcontents

\section{Introduction}
Various scientific experiments in metrology, radio astronomy or particle accelerators require the syntonization or synchronization between remotely located sites \cite{Cliche2006, Chou2010a, Gumerlock2014}. Also applications like telecommunication and navigation rely on precise synchronization among remote frequency sources \cite{Cacciapuoti2009}. To take advantage of the rapid increase in performance of atomic clocks which have recently been reported to achieve an instability and accuracy at a level of $1~\times~10^{-18}$ \cite{Bloom2014, Ushijima2014}, novel frequency dissemination techniques capable of supporting the performance of state-of-the-art clocks are being developed. In recent years, extensive research on the transfer of stable optical frequencies via optical fiber links demonstrated excellent performances with residual instabilities of a few parts in $10^{19}$ \cite{Droste2013, Bercy2014, Calonico2014a}. This method, however, requires a fiber link connection between the remote sites which might be impractical for some geographical regions or certain applications. Additionally, the establishment of intercontinental optical fiber links for frequency dissemination will be challenging.

A more traditional way of transmitting time or frequency information is based on exchanging microwave signals between ground stations and satellites. Here, two existing techniques have to be distinguished. If a geostationary telecommunication satellite is used as a space based repeater station, microwave signals are exchanged between two remote locations on the earth. In this approach, signals are sent from and received by both locations simultaneously in order to cancel out most one-way propagation delay effects. This method is typically referred to as Two-Way Satellite Time and Frequency Transfer (TWSTFT) and requires complex equipment and costly transponder capacity on the commercial satellites \cite{Kirchner1991, Bauch2011, Fujieda2012}. An alternative method is based on Global Navigation Satellite Systems (GNSS) such as the Global Positioning System (GPS) to remotely synchronize frequency standards by simply receiving the signals transmitted from the satellites \cite{Levine2008}. Because of its simplicity and cost efficiency this method is used by most metrology institutes and timing laboratories to compare the majority of atomic frequency standards worldwide.

A recent comparison between a TWSTFT and a GPS carrier-phase (CP) link over a baseline of 9,000~km revealed a frequency difference of up to $9.5~\times~10^{-16}$ between the two methods which exceeded the estimated statistical uncertainty \cite{Fujieda2014}. In our current study, we aim to assess the frequency transfer capabilities of a GPS link based on a state-of-the-art Precise Point Positioning analysis over a baseline of 450~km. We employ a 920~km phase-stabilized optical fiber link \cite{Predehl2012}, which serves as a reference link to transfer frequency information between the two endpoints of the GPS link with very low uncertainty.

The consistency of the results achieved independently via satellite transfer and fiber transfer provides an upper limit for the accuracy and instability of each of the transfer techniques.

\section{Methods and experimental setup}
Nowadays, the comparison of atomic frequency standards, for example hydrogen masers, is straightforward. In the simplest case, an antenna capable of receiving signals that are broadcasted by the constellation of GPS satellites and a suitable receiver are used to derive the difference between the phase of the incoming signals and the local frequency standard connected to the receiver \cite{Levine2008}. Recording the phase difference between the transmitted GPS signals and the two frequency standards simultaneously generates two sets of data that can be used to calculate the phase difference between the two frequency standards.

The frequency transfer capabilities of such a GPS link have been shown to be commensurate to the frequency instability of the active hydrogen masers under comparison \cite{Bauch2006, Fujieda2012, Piester2009}. In a comparison of two such masers it is therefore challenging to separate the individual contributions from each maser and from the GPS link itself. Even though it is expected that the noise of a GPS link dominates a maser comparison for short averaging times, it remains unclear at which level the noise of the masers start to dominate and subsequently what the ultimate performance supported by a GPS CP link is, that could be used in case superior frequency standards would be available.

In contrast to that, the frequency instability achieved when transferring a stable optical frequency signal along the phase-stabilized 920~km fiber link between the Max-Planck-Institut für Quantenoptik (MPQ) and the Physikalisch-Technische Bundesanstalt (PTB) \cite{Predehl2012} is at least two orders of magnitude below that of an active hydrogen maser for any relevant averaging time. We intend to use the superior performance of this fiber link to circumvent the afore mentioned issue of non-separable noise sources by eliminating the contribution of the masers.

\begin{figure}[t]
\centering
\includegraphics[width=0.8\textwidth]{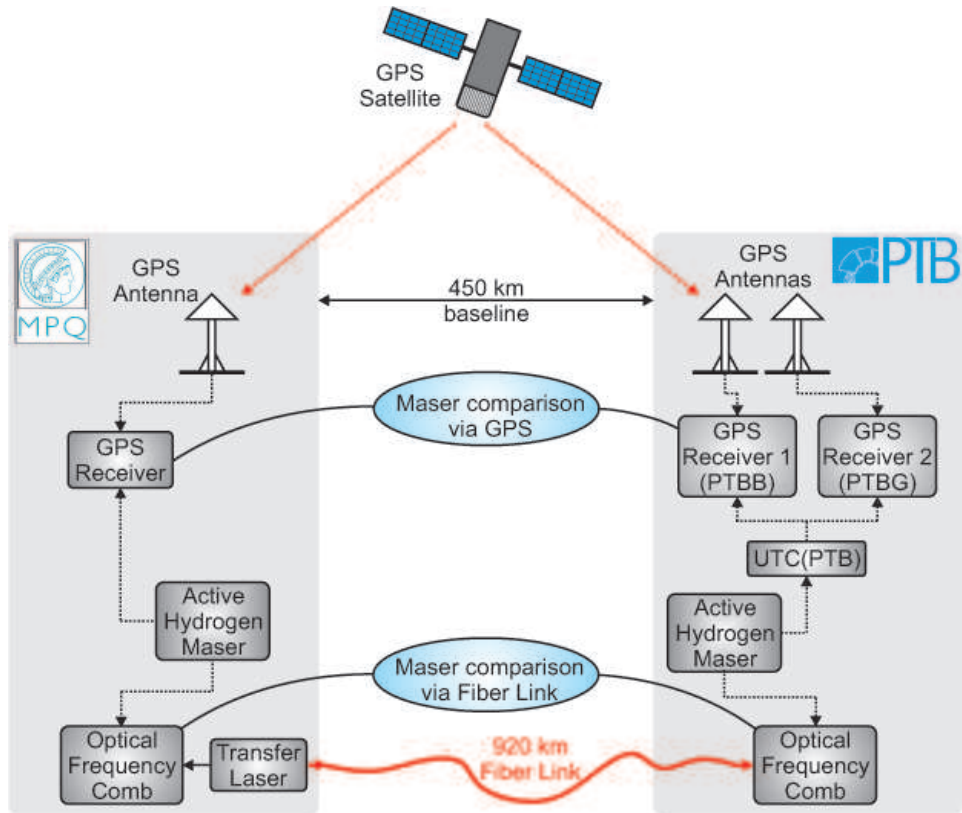}
\caption[Transmission scheme for the GPS-CP links characterization]{Experimental setup for the characterization of the GPS link between MPQ and PTB. Two hydrogen masers are compared via a 920~km fiber link and via a GPS link simultaneously. At each site, an optical frequency comb is referenced to the local maser. The fiber link is operated from MPQ and the maser comparison via fiber link is accomplished by measuring the transfer laser frequency against the optical frequency combs. The maser comparison via GPS is performed by measuring the maser frequency against the GPS signal.}
\label{fig:Fig1}
\end{figure}

\Fref{fig:Fig1} illustrates the experimental setup. The GPS link to be characterized has been operated between the two institutes MPQ and PTB while simultaneously performing an optical frequency transfer via the 920~km fiber link. Active hydrogen masers are operated at each institute which are separated by a geodesic distance of about 450~km. The characterization of the GPS link is accomplished by comparing those two masers over the fiber link and over the GPS link simultaneously. In the difference between the comparison over the GPS link and over the fiber link (the double difference) the noise contributions of the masers drop out. Due to the superior performance of the fiber link, the resulting double difference solely reflects the instability of the GPS link.

The maser comparison via the fiber link is realized by transferring a highly stable optical frequency from MPQ to PTB. The fiber link introduces noise to the optical signal due to environmental perturbations that have to be compensated by an interferometric noise cancellation system which is operated at MPQ \cite{Predehl2012}. Optical frequency combs (Menlo Systems GmbH) at both institutes are referenced to the local masers and connect the optical and microwave frequencies. In the fiber link system, we generate heterodyne beat signals by superimposing two laser beams on a photo detector in order to stabilize the link transfer and to measure the optical frequency to be transferred. All optical heterodyne beat signals in the fiber link itself as well as in the frequency combs are counted with high-resolution frequency counters ($\Lambda$-type, K~+~K Messtechnik GmbH) synchronized between MPQ and PTB. They are operated with a gate time of 1~s. The GPS link is established using commercially available GPS receivers. At MPQ, a GPS receiver (Septentrio PolaRx2e) is used which gets its internal frequency reference via a 10~MHz signal from the maser operated at MPQ. Data from two different GPS receivers operated at PTB (both Ashtech Z-XII3T) are used which permits an additional comparison of the two different receivers among one another. Both receivers are connected to 10~MHz and 1~PPS signals representing PTB's reference time scale UTC(PTB) \cite{Bauch2012} as they constitute the pivot point for all GPS-based time comparisons made worldwide in the context of the realization of Coordinated Universal Time (UTC) by the International Bureau of Weight and Measures (BIPM) \cite{Arias2011}. The 10~MHz signal from the maser at PTB, connected to the local frequency comb, is thus measured against the UTC(PTB) frequency signal with the help of a phase comparator (Timetech PCO 10265).

During recent years it has become more and more common to build on GPS-based frequency comparison techniques that were initially developed for positioning. Precise Point Positioning (PPP), for instance, is a technique providing position with a high accuracy on a global scale with a single isolated (not part of a network) GNSS receiver in post-processing. It uses code and carrier-phase measurements that are collected in geodetic GPS receivers. Instead of differencing observations made at various sites, PPP builds on the precise satellite orbit, clock products and troposphere parameters generated by the International GNSS Service (IGS) \cite{Dow2009}. Different software packages for the PPP analysis of GPS data are available. They differ in the details of the algorithmic combination of observations. In our study we use the software package provided by Natural Resources Canada (NRC) that was made generously available free of charge to several timing laboratories \cite{Kouba2001}. Nowadays, this software is used regularly by BIPM to calculate PPP-based frequency comparisons among major international timing institutes as part of the realization of UTC. The NRCan software allows the processing of periods in excess of one day so that day boundary jumps are avoided and the GPS data shown in \fref{fig:Fig5} are processed in one run. Note that the apparent gaps in the GPS data result from the unavailability of the optical fiber link data. \Fref{fig:Fig5} only shows periods when both links were operational.

The use of PPP appears particularly attractive for the current study as it adapts to a global but sparse network of stations. MPQ represents a station equipped with a high-quality local frequency reference, but it is separated from the network traditionally cooperating with the BIPM. The timing results provided by the NRCan-PPP software represents the time difference between the local clock and IGS time. IGS time is generated as an average of a subset of atomic clocks (in particular active hydrogen masers) of stations affiliated with the IGS. IGS time is loosely steered towards GPS time \cite{Senior2003}. GPS time on the other hand is the internal reference time scale of GPS and is used in the transmitted GPS signal for reporting the individual satellite clock signal to the user. GPS time is a time scale composed of ground clocks and some satellite clocks and is steered towards UTC(USNO), the realization of UTC of the United States Naval Observatory. The data from the MPQ and PTB receivers are processed with the NRCan software package using IGS orbit and IGS 30~s clock products \cite{Kouba2001}. The sample rate is chosen to 30~s at MPQ and PTB.

\section{Signal validation and uncertainty contributions}

\begin{figure}[t]
\includegraphics[width=0.8\textwidth]{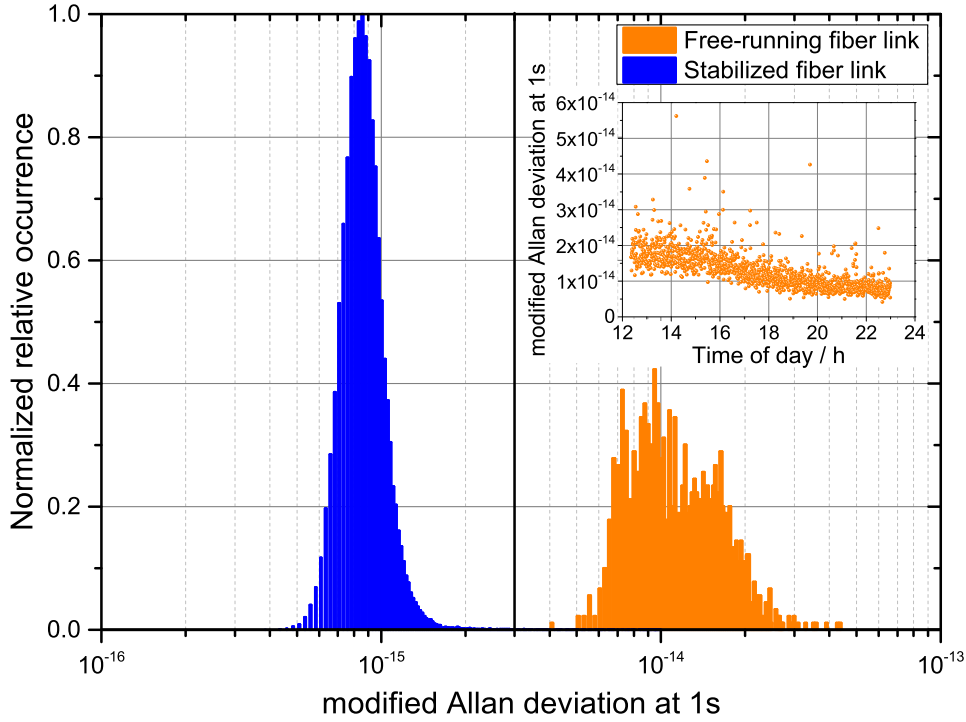}
\caption[Verification of the fiber link operation]{1-s modified Allan deviation determined from 30 individual adjacent frequency measurements of heterodyne beat notes between the transferred light from MPQ and two stable optical references at PTB. The data shown was determined from a 30 day measurement campaign. If the fiber link transfer is stabilized, $\sigma_y(1s)$ is about $9~\times~10^{-16}$ while it increases to $\approx~1~\times~10^{-14}$ if the stabilization is deactivated. A threshold of $3~\times~10^{-15}$ is introduced to verify a proper fiber link operation. The inlet shows $\sigma_y(1s)$ over the time of the day, indicating a noise reduction during the night as observed in a previous study \cite{Droste2013}.}
\label{fig:Fig2}
\end{figure}

The functionality of the fiber link is verified by calculating the instability of the transferred frequency at PTB against two stable optical references. If the fiber induced noise cancellation is deactivated, these heterodyne beat notes show a 1-s instability of about $1~\times~10^{-14}$ as shown in \fref{fig:Fig2}. When the noise cancellation control loop is active on the other hand, the 1-s instability decreases to about $9~\times~10^{-16}$ so that this measure can be used to monitor the operation of the fiber links active stabilization. The 1-s instability of these two beat signals was determined from 30 individual adjacent frequency measurements. If the 1-s instability exceeds a threshold of $3~\times~10^{-15}$ for both signals, we discard all of those 30 data points. To detect cycle-slips in the optical part of the system which includes the frequency combs, we apply a redundant counting scheme in analogy to previous experiments \cite{Predehl2012, Droste2013}. All data points for which the two redundant counted signals disagree by more than a predefined threshold are discarded to prevent them from entering the data analysis. This threshold is adapted to the noise of the individual signals by calculating the medium absolute deviation (MAD). We find a robust value for the cycle-slip threshold to be 8~$\times$~MAD in the sense that varying this threshold did not change the amount of detected cycle-slips significantly.

Due to the different sampling intervals of the fiber and GPS link data (1~s versus 30~s), the combination of both data sets requires some preprocessing of the fiber link data. The most intuitive approach is to average 30 1-s fiber-link-data-points to equalize the sampling intervals. However, one single cycle-slip in the fiber link data would lead to a rejection of the remaining 29 data points within the corresponding GPS data window. However, the instability contribution of the maser comparison over the optical link can be neglected as long as each 30~s interval contains at least 10 valid data points of the optical transfer: The frequency difference between the two active hydrogen masers measured over the fiber link shows an instability of $\approx~1~\times~10^{-13}$ in 1~s as no excess noise is introduced by the fiber link. The frequency difference of the masers measured over the GPS link, however, has an instability of $\approx~3~\times~10^{-13}$ in 30~s. In the worst case, all 10 fiber link data points will be incoherent (i.e. non-contiguous) due to cycle-slips which results in an instability of $1~\times~10^{-13}/\sqrt{10}~\approx~3~\times~10^{-14}$. Therefore, the instability of the $\ge$~10-s averaged fiber link data points will always be at least one order of magnitude below that of the GPS link data points. After applying a non-weighted average to the fiber link data we subtract the fiber link data from the GPS link data. This results in the double difference which reveals the GPS link performance without the contributions from the masers.

\begin{figure}[b]
\includegraphics[width=0.8\textwidth]{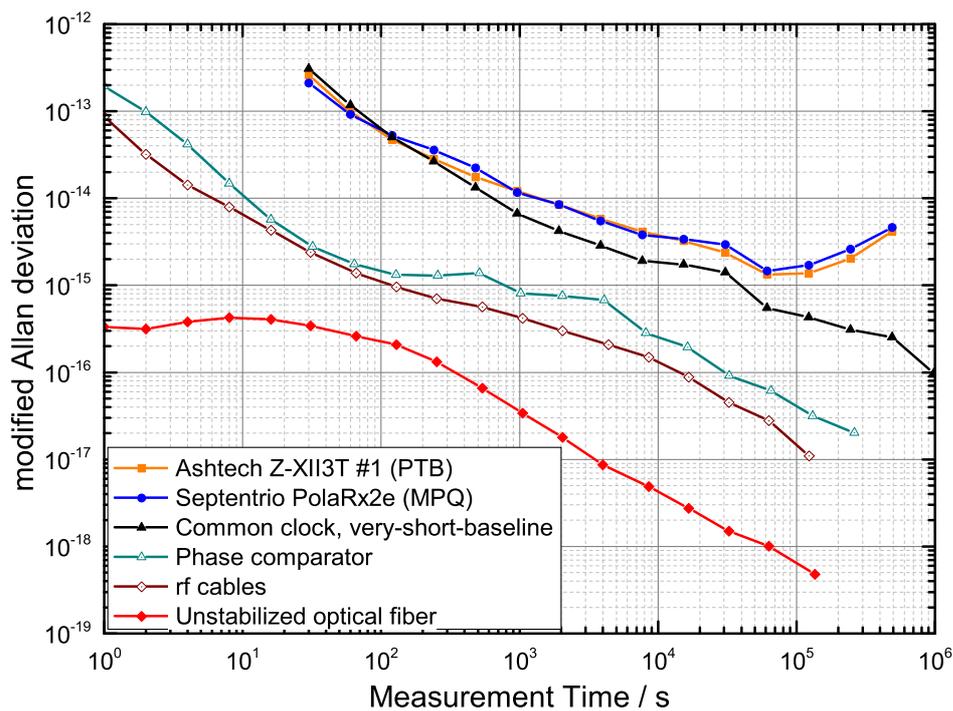}
\caption[Comparison of GPS receivers]{Fractional frequency instability of the difference between IGS time and the hydrogen masers for the receiver at MPQ (filled blue circles) and one of the receivers at PTB (filled orange squares). The common clock very-short-baseline realized between the two setups at PTB (filled black triangles) provides a measure of the noise floor for this kind of data analysis. Contributions from other components in the system like the phase comparator (open green triangles), from rf cables (open brown diamonds) and from unstabilized optical fibers (filled red diamonds) are well below the instability determined from the common clock very-short-baseline configuration.}
\label{fig:Fig3}
\end{figure}

We identified and experimentally studied the uncertainty and instability contributions of several components in the system, summarized in \fref{fig:Fig3}. The three GNSS receivers used here represent the state of the art in geodetic and timing applications. The PPP data analysis provides the phase difference between IGS time and the masers connected to the receivers using the carrier-phase observables via a linear combination at the two GPS frequencies L1 and L2 that removes the first order disturbance due to the signal propagation through the ionosphere \cite{Kouba2001}. The frequency instability expressed as the modified Allan deviation is shown in \fref{fig:Fig3} for the receiver at MPQ and for one of the receivers at PTB. The data represent the combined instability of the masers, IGS time and the contributions of signal propagation and processing. 

Each PTB receiver is connected via an about 50~m long cable to its individual antenna. The two antennas are separated by only a few meters on the roof of the PTB building. Since the two receivers are connected to the same maser, this part of the setup constitutes a common clock very-short-baseline configuration, which is analyzed using the NRCan PPP software. Similar investigations have been performed in \cite{Defraigne2011, Petit2015}. Such a comparison is not affected by the frequency instability of the masers, the effects of signal propagation through the ionosphere and instabilities of the IGS time. Each of the two receivers involved, however, are equipped with separate antennas that may be affected by multipath propagation in a slightly different way, and also the signal processing in the two receivers follows different algorithms. Such effects lead to an unavoidable noise floor in the comparisons. Additionally, the PPP software estimates troposphere parameters independently although in principle the propagation conditions should be equal for both closely located antennas. Non-standard software would be necessary to avoid the noise contribution related to this.

It can be seen in \fref{fig:Fig3} that in our case an instability of about $1~\times~10^{-16}$ is reached after an averaging time of $10^6$~s. The mean frequency difference was measured to $2.5~\times~10^{-17}$, thereby excluding a significant systematic error.

The phase comparator used at PTB may also constitute a limiting factor. It is known that the device may produce a measurement error that depends on the frequency difference between the two signals that are being compared. Therefore, UTC(PTB) and maser signals were compared in two different types of phase comparators simultaneously. From the difference of the two phase comparator outputs we derive an upper limit for the contributions to the measurement instability and uncertainty. In \fref{fig:Fig3} it is shown that the contribution of the phase comparator to the frequency instability is below the one of the common clock at all relevant measurement times. As the relative mean difference of the two phase comparator results is about $1~\times~10^{-18}$, a significant uncertainty contribution can be excluded.

In \fref{fig:Fig1} a connection between the maser and the frequency comb is sketched that actually represents a 185~m long rf cable connecting two buildings. The measured frequency instability for a signal transferred through such a cable is shown in \fref{fig:Fig3} as open diamonds. The contribution from this cable is about one order of magnitude below that of the common clock for all measurement times and the mean frequency is determined to $3.6~\times~10^{-17}$ and does therefore not constitute a significant source of error.

Since optical fibers are sensitive to environmental perturbations, unstabilized fiber sections might introduce a significant amount of noise to the signals. The longest unstabilized fiber section in our setup is about 11~m long. The contribution from this fiber is shown in \fref{fig:Fig3} as filled diamonds. With a relative mean of $3~\times~10^{-19}$, the contribution from this fiber is negligible.

Thus, the investigated components revealed no systematic shifts within the statistical uncertainty derived from the instability. \Fref{fig:Fig3} and the measured mean frequencies indicate that the dominant source of instability and uncertainty of the maser comparison will be linked to the GPS comparison itself. Increasing the baseline from a few meters to 450~km will add additional noise as the signals from the satellites pass through different atmospheric sections. In the following, we aim to determine the influence of the longer baseline and to quantify the uncertainty that is associated with such a GPS CP link.

\section{Results}

\begin{figure}[b]
\centering
\includegraphics[width=0.8\textwidth]{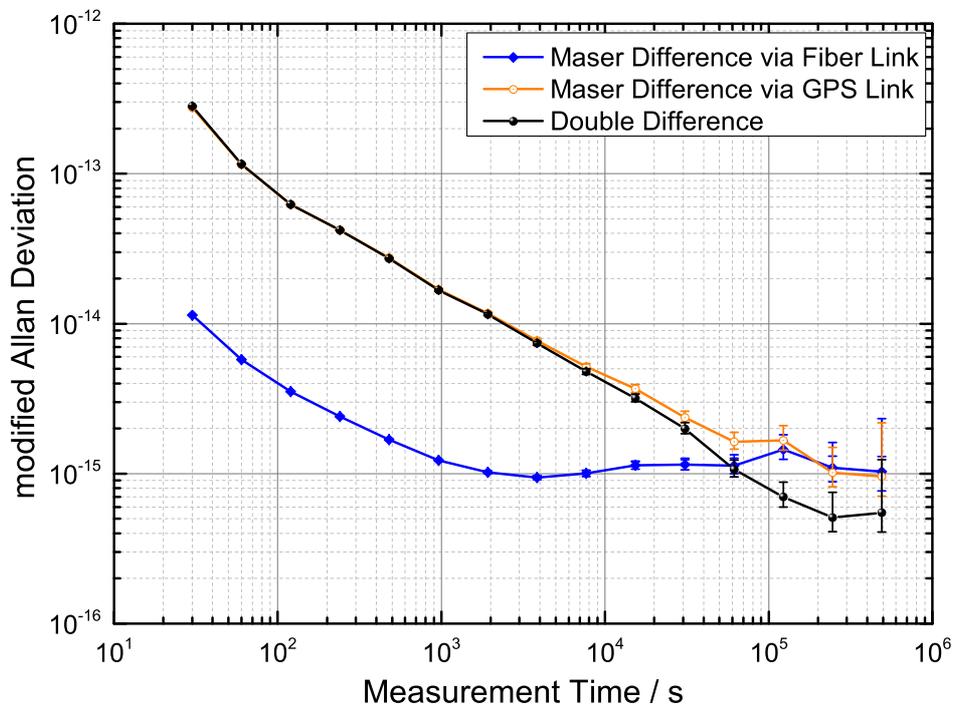}
\caption[Frequency instability of maser difference]{Frequency instability of the maser difference measured via the fiber link and via the GPS link, respectively. The double difference reveals the true GPS link performance without any contribution from the masers.}
\label{fig:Fig4}
\end{figure}

The operation of two frequency transfer links in both the microwave and the optical domain simultaneously involves a large amount of scientific equipment. The proper operation of every component has to be verified as well as the connection between the two links and frequency domains. Due to the complexity of the system, we first conducted a test measurement over the course of a few weeks in January 2014. The insights gained in this first campaign are used in an extended measurement campaign with a duration of approximately four weeks, lasting from 4 April to 4 May 2014. The results of the first test measurement are in good agreement with the results of the extended campaign discussed below. We measure the difference of the two masers via the two links and calculate the frequency instability of these signals which is shown in \fref{fig:Fig4}.

The comparison via the fiber link indicates the difference of the masers practically without any noise contribution from the optical transfer. In contrast to that, the maser comparison via the GPS link is dominated by noise components from the GPS link itself, at least for short averaging times. For long measurement times the instabilities of GPS transfer and optical transfer become comparable as the noise contribution of the masers becomes dominant. Forming double differences suppresses the maser noise to a high extent and therefore reflects the true GPS link instability in good approximation. In order to get perfect noise suppression, the measurement intervals during which the optical link and the GPS link data are collected need to be exactly identical. The instability of the double difference averages down to a level of $5~\times~10^{-16}$ after 500,000~s. This is close to the value of $3~\times~10^{-16}$ measured in the common clock configuration (see \fref{fig:Fig3}) for the same measurement time. The double difference in \fref{fig:Fig4} raises the question whether we reach a noise floor for measurement times >~200,000~s. If we form one continuous data set by merging the data of the test measurement (in January 2014) and the April campaign, we can calculate instability values for even longer measurement times, to further search for such a noise floor in the double difference. We find the frequency instability actually drops to about $3~\times~10^{-16}$ at 600,000~s. 

\begin{figure}[b]
\centering
\includegraphics[width=0.8\textwidth]{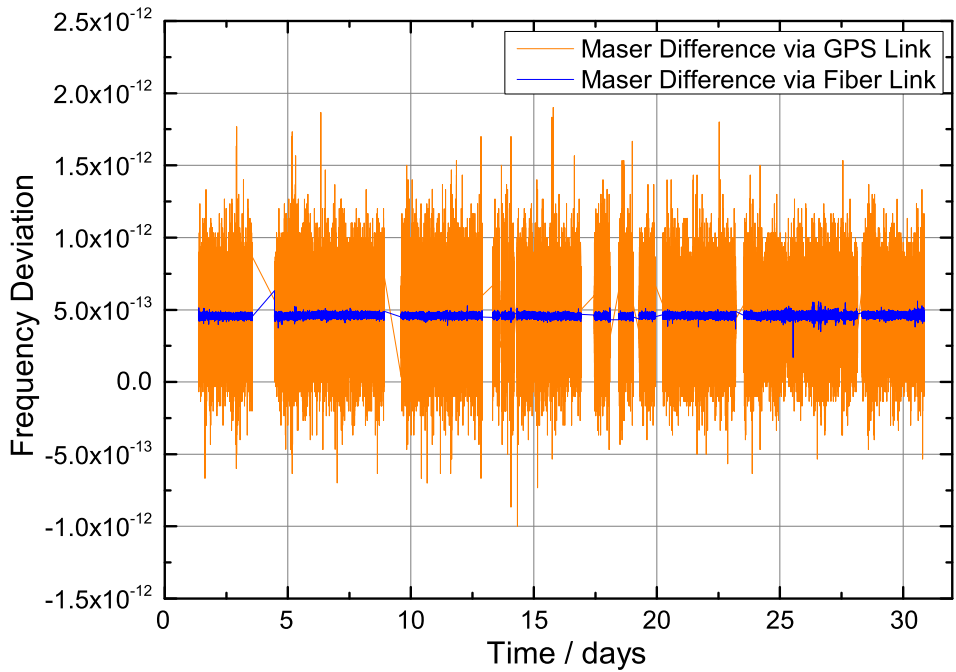}
\caption[Frequency deviation of masers]{Frequency deviation between the two masers at MPQ and PTB measured over the GPS link and over the fiber link, respectively. The masers show a mean frequency difference of about $4.6~\times~10^{-13}$ with respect to each other. The occasional gaps in the fiber link data are due to cycle-slips and a malfunction of one of the frequency comb systems. GPS link data are only shown at intervals when the fiber link data were available.}
\label{fig:Fig5}
\end{figure}

The accuracy of the fiber link has been constrained to a few parts in $10^{19}$ \cite{Predehl2012} so that any deviation between the two comparisons greater than this value can be attributed to the GPS link. \Fref{fig:Fig5} shows the frequency deviation between the masers measured via the GPS link and via the fiber link. The gaps in the trace result from cycle-slips in the fiber link data as well as from a malfunction of one of the frequency comb systems. 

\begin{table}
\centering 
\begin{tabular}{p{5.5cm}cc}\hline
	 \rule{0pt}{2.5ex}Measured signal & Arithmetic mean & Statistical uncertainty\\ \hline \hline
   \rule{0pt}{3ex}Maser difference via fiber link & $4.595~\times~10^{-13}$ & N/A\\ \noalign{\vskip 1mm}
	 Maser difference via GPS link & $4.597~\times~10^{-13}$ & N/A\\ \noalign{\vskip 1mm}
	 Double difference & $2.1~\times~10^{-16}$ & $6.0~\times~10^{-16}$\\ \noalign{\vskip 1mm}\hline
\end{tabular}
\caption[Results of the maser difference]{Results of the maser difference measured via the fiber and via the GPS link together with the results of the double difference. The results of the double difference are calculated from 300~s data as explained in the text.}
\label{tab:Table1}
\end{table}

The arithmetic mean of the maser difference measured via the fiber link and via the GPS link are shown in \tref{tab:Table1}. The results are obtained from a total of 71,375 data points where each data point represents a measurement interval of 30~s. We calculate the double difference by subtracting the two data sets of \fref{fig:Fig5} in order to eliminate the contributions of the masers. The statistical uncertainty ($\sigma/\sqrt{N}$ where $\sigma$ is the standard deviation and $N$ the number of data points) of the double difference given in \tref{tab:Table1} is limited by the GPS link data. We apply a non-weighted average to the GPS link data by combining 10 GPS data points, thus representing a measurement time of 300~s (see \cite{Predehl2012, Droste2013} for details). In analogy to the procedure described above, we select and average a minimum of 100 individual fiber link data points that lay within the new 300~s GPS measurement window (thus $N=7137$). In the resulting double difference we can constrain any offset between the two frequency transfer methods to $(2.11~\pm~5.97)~\times~10^{-16}$. The overall mean frequency for the joint data set (January and April campaigns) is $1.3~\times~10^{-16}$.

\begin{figure}[t]
\centering
\includegraphics[width=0.8\textwidth]{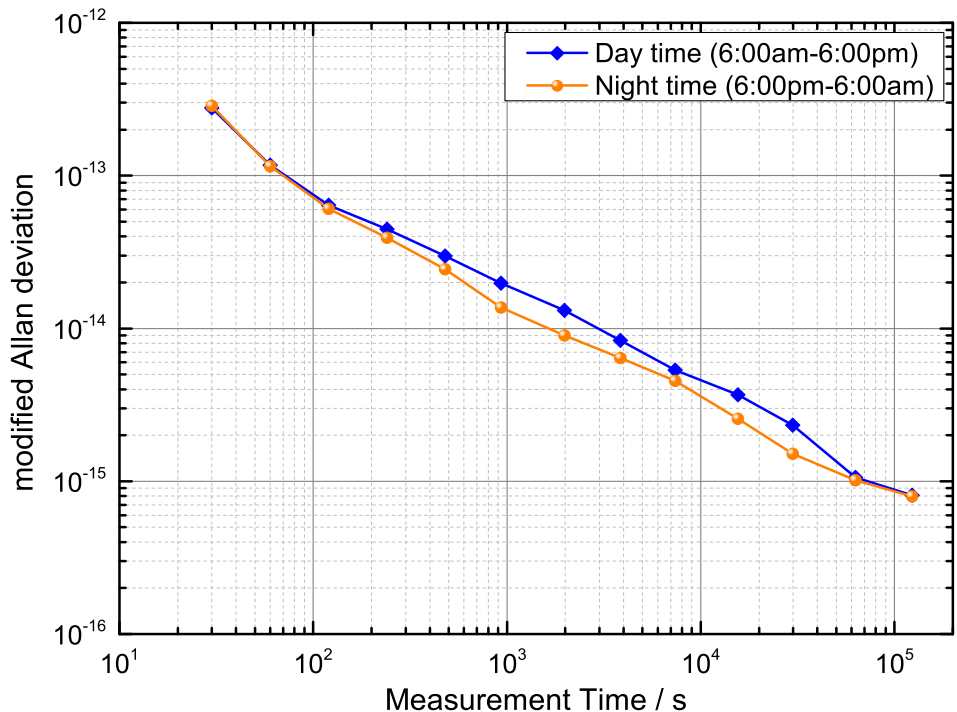}
\caption[Frequency instability of the GPS link at day and night]{Frequency instability of the double difference for the day time from 6:00~am to 6:00~pm and for the night time from 6:00~pm to 6:00~am.}
\label{fig:Fig6}
\end{figure}

It is of interest whether the GPS link frequency transfer shows diurnal variations. We separate our measurement data into day time and night time (cut-off at 6am/6pm). \Fref{fig:Fig6} shows the frequency instability of the double difference at day and at night. The difference between the day and night frequency instability is below a factor 1.6 at all measurement times. The mean frequency for day and night data was identical within the measurement uncertainty.

The difference in height above the earth geoid between MPQ and PTB is about 400~m, corresponding to a gravitational redshift of $4.4~\times~10^{-14}$. However, this effect does not have to be taken into account as it cancels in the double difference between the two maser comparisons.

\section{Discussions}
The values stated above are determined by averaging over roughly 600 hours of valid data measured over the period of about one month. When we compare the two masers via the optical link, see the blue curve in \fref{fig:Fig4}, the instability is about $1~\times~10^{-14}$ in 30~s. The instability of $3~\times~10^{-13}$ in 30~s observed in the GPS link comparison is clearly limited by the noise of the GPS link transfer. At about $10^4$~s the noise of the masers themselves starts to visibly contribute to the comparison via the GPS link as can be seen by the splitting of the curves for the GPS link and the double difference in which the contributions of the masers drop out. The comparison results of UTC(PTB) and UTC(OP) via a PPP GPS link during 2014 are available at the BIPM ftp server \cite{BIPM2014} and the results fit very well to the data for the PTB to MPQ link.

The experiment presented here is the first point-to-point frequency comparison between two independent frequency sources via an optical fiber of such length. Loop experiments gave evidence that the frequency transfer accuracy of such a fiber-based system is excellent, nevertheless it is justified from a metrological point of view to look for an independent assessment of the performance. Here, the GPS PPP link is the best affordable alternative and it provides at least an upper limit for the achieved accuracy of the fiber-based frequency comparison. In the near future, comparisons of optical frequency standards with uncertainties below $10^{-17}$ will surely better serve the purpose.

\section{Conclusions}
We characterized a GPS CP frequency transfer link by comparing two hydrogen masers that are separated by a physical distance of 450~km over a GPS link and over a phase-stabilized optical fiber link. A short-term instability of the GPS link of $3~\times~10^{-13}$ in 30~s was observed. The parallel operation of a GPS link and a fiber link allowed us to characterize the GPS transfer on timescales of weeks without the contribution of the local oscillators (hydrogen masers). We demonstrated that a GPS CP link ultimately supports an instability and accuracy of below $6~\times~10^{-16}$. We exclusively operated standard commercially available equipment for the GPS link and processed all observations from the GPS receivers with the commonly used NRCan PPP software. Note, that very recent investigations also show improvements on the PPP software \cite{Petit2015}.

\ack
We acknowledge financial support by the SFB-1128 geo-Q on "Relativistic Geodesy and Gravimetry with Quantum Sensors", and the European Metrology Research Programme (EMRP) under SIB-02 NEAT-FT and SIB-60 Surveying. The EMRP is jointly funded by the EMRP participating countries within EURAMET and the European Union. We thank the members of Deutsches Forschungsnetz in Berlin, Leipzig, and Erlangen, Germany, as well as Gasline GmbH for a fruitful collaboration. PTB acknowledges Natural Resources Canada for granting the license of the PPP software package.

\section*{References}

\bibliographystyle{iopart-num}

\providecommand{\newblock}{}

\end{document}